\documentclass{aa}

\usepackage{txfonts}

\usepackage{graphicx}
\usepackage[colorlinks]{hyperref}
\usepackage{xcolor}

\hypersetup{
    linkcolor=blue,
    citecolor=blue,
    filecolor=magenta,      
    urlcolor=blue
}

\newcommand{\Msun}{\mbox{$\mathrm{M}_{\odot}$}}

\newcommand{\Rsun}{\mbox{$\mathrm{R}_{\odot}$}}

\newcommand{\Mi}{\mbox{$M_\mathrm{i}$}}
\newcommand{\omegai}{\mbox{$\omega_\mathrm{i}$}}

\newcommand{\Teff}{\mbox{$T_{\mathrm{eff}}$}}

\defcitealias{2022A&A...665A.126N}{Paper I}
\newcommand{\paperone}{\citetalias{2022A&A...665A.126N}\xspace}

\begin{document}

\title{PARSEC V2.0: Rotating tracks and isochrones for seven additional metallicities in the range $Z=0.0001-0.03$ \thanks{All of our models are released at two dedicated websites: \url{https://stev.oapd.inaf.it/PARSEC/tracks_database.html} for stellar tracks, and \url{https://stev.oapd.inaf.it/cgi-bin/cmd_3.8} for isochrones.}}
\titlerunning{\texttt{PARSEC V2.0} tracks and isochrones}

\titlerunning{\texttt{PARSEC V2.0} tracks and isochrones}
\subtitle{}

   \author{C. T. Nguyen
          \inst{1,2,3}
          \and
          G. Costa       
          \inst{4, 5}
          \and
          A. Bressan
          \inst{3}
          \and
          L. Girardi    
          \inst{5}
          \and
          G. Cescutti          
          \inst{1,2}
          \and
          A.J. Korn          
          \inst{6}
          \and
          G. Volpato  
         \inst{7,5}
         \and
          Y. Chen
         \inst{8}
         \and
          G. Pastorelli
         \inst{4}
         \and 
          M. Trabucchi
          \inst{4,5}
         \and 
          K. G. Shepherd
          \inst{3,5}
         \and
          G. Ettorre  
          \inst{4,9,10}
         \and
          S. Zaggia 
          \inst{5}
          }

        \institute{
        INAF Osservatorio Astronomico di Trieste, Via Giambattista Tiepolo, 11, Trieste, Italy,\\
        \email{chi.nguyen@inaf.it}
        \and
        University of Trieste, Piazzale Europa, 1, Trieste, Italy
        \and
        SISSA, Via Bonomea 265, I-34136 Trieste, Italy,
        \and
        Dipartimento di Fisica e Astronomia, Universit\`a degli studi di Padova,
        Vicolo dell'Osservatorio 3, Padova, Italy
        \and
        INAF - Osservatorio Astronomico di Padova, Vicolo dell'Osservatorio 5, Padova, Italy
        \and
        Department of Physics and Astronomy, Uppsala University, Box 516, SE-75120 Uppsala, Sweden
        \and
        Institut d'Astronomie et d'Astrophysique, Université Libre de Bruxelles, 226-Boulevards du Triomphe-B 1050 Bruxelles, Belgium
        \and
        Anhui University, Hefei 230601, China
        \and
        Dipartimento di Fisica e Astronomia, Universita` di Bologna, Via Gobetti 93/2, I-40129 Bologna, Italy
        \and
        INAF – Astrophysics and Space Science Observatory of Bologna, Via Gobetti 93/3, 40129 Bologna, Italy
      }

\authorrunning{Nguyen et al.}

   \date{}

\hypersetup{
    linkcolor=blue,
    citecolor=blue,
    filecolor=magenta,      
    urlcolor=blue
}

\abstract{
\texttt{PARSEC v2.0} rotating stellar tracks were previously presented for six metallicity values from subsolar to solar values, with initial rotation rates ($\omegai$, defined as the ratio of the angular velocity and its critical value) spanning from the non-rotating case to very near the critical velocity (i.e. $\omegai=0.99$), and for initial masses covering the $\sim 0.7~\Msun$ to $14~\Msun$ interval. Furthermore, we provided the corresponding isochrones converted into several photometric systems for different inclination angles between the line-of-sight and the rotation axes, from $0^\circ$ (pole-on) to $90^\circ$ (equator-on). 
In this work, we expand this database with seven other metallicity sets, including five sets of low metallicity ($Z=0.0001-0.002$) and two sets of super-solar values (up to $Z=0.03$). 
We present the new stellar tracks, which comprise $\sim$3\,040 tracks in total ($\sim$5\,500 including previous sets), along with the new corresponding rotating isochrones. We also introduce the possibility of creating isochrones by interpolation for values of rotating rates that were not available in the initial set of tracks. 
We compared a selection of our new models with rotating stellar tracks from the Geneva Stellar Evolution Code, and we assessed the quality of our new tracks by fitting the colour-magnitude diagram of the open cluster NGC~6067. We took advantage of the projected rotational velocity of member stars measured by Gaia to validate our results and examined the surface oxygen abundances in comparison with the observed data. 
All newly computed stellar tracks and isochrones can be retrieved via our dedicated web databases and interfaces. 
}

   
   \maketitle
   
\section{Introduction}

Rotation is known to play an important role in stellar structure and evolution. 
It induces two main effects, which are the departure from the spherical shape and the enhancement of mixing processes. The first effect is due to the centrifugal forces that can strongly geometrically distort the stellar surface, which directly affects the photometry of stars and their position in the colour-magnitude diagram (CMD). This aspect was studied by several authors \citep[see][and references therein]{2009MNRAS.398L..11B, 2013ApJ...776..112Y, 2019MNRAS.488..696G}. It was claimed that it might explain some peculiar properties of young and intermediate-age stellar clusters, such as the main-sequence (MS) split \citep[][]{2017MNRAS.465.4363M,2019A&A...631A.128C,2022ApJ...938...42H} and the extended MS turn-off \citep[][]{2009MNRAS.398L..11B, brandt15, 2018MNRAS.477.2640M,2018ApJ...869..139C}.
The second important effect is due to rotational instabilities such as meridional circulation and shear instability \citep[see][]{2009pfer.book.....M,2000A&A...361..101M,2013ApJ...764...21C}, which transport chemical elements across the stable radiative layers of the stars. This additional mixing enriches the stellar surface with processed elements and provides fresh fuel to the core, which in turn increases the MS lifetimes and thus the core mass and luminosity in the post-MS phases.

These significant rotation effects have been included in some libraries of stellar evolutionary tracks, such as those derived from the evolution codes: Geneva stellar evolution \citep[\texttt{GENEC},][]{2012A&A...537A.146E,georgy12,2013A&A...558A.103G,groh19,yusof22,2024A&A...690A..91S}, Montpellier-Geneva \citep[\texttt{STAREVOL},][]{2019A&A...631A..77A,borisov24}, Modules for Experiments in Stellar Astrophysics \citep[\texttt{MESA},][]{paxton19,hastings23}, Frascati Raphson Newton Evolutionary Code \citep[\texttt{FRANEC},][]{limongi18} and Bonn \citep{2011A&A...530A.115B}. 
Rotation was also introduced in the PAdova and tRieste Stellar Evolutionary Code (\texttt{PARSEC}) in its version 2.0 \citep[][]{2019A&A...631A.128C,2019MNRAS.485.4641C,2019MNRAS.488..696G}. A large grid of \texttt{PARSEC} evolutionary tracks and isochrones for rotating stars was presented by \citet[][hereafter \paperone]{2022A&A...665A.126N}, comprising six values of the initial metallicity ($Z$, in mass fraction) from 0.004 to 0.017. These tracks were released together with the corresponding isochrones in many different photometric systems for seven values of the initial rotation rates $\omegai$. We recall that $\omegai$ is defined as the surface angular velocity divided by the critical angular break-up velocity at which the centrifugal forces cause the equatorial layers to become detached from the star.
In this work, our goal is to expand the \texttt{PARSEC v2.0} database with seven more sets of metallicity. The database then covers the range from $Z=0.0001$ to $0.03$. 
Moreover, an improved interpolation method \citep[already used in][]{2025MNRAS.tmp..559E} allows us to interpolate between any two values of $\omegai$. Therefore, we can now provide isochrones with any $\omegai$ in the interval $0.00-0.99$. 

This paper is organised as follows. Sect.~\ref{sec:inputphysics} summarises the input physics we adopted to compute the stellar tracks. Sects.~\ref{sec:tracks} and \ref{sec:isochrones} illustrate the tracks and isochrones that were newly computed. Sect.~\ref{sec:summary} compares our results with the observed data of the open cluster NGC~6067 and draws some final conclusions.

\section{Input physics}
\label{sec:inputphysics}

The input physics of our models is as described in Section 2 of \paperone. For the sake of clarity, we summarise the most important ingredients in this section. 

Rotation was first implemented in \texttt{PARSEC} by \citet{2019MNRAS.485.4641C}. 
They investigated the concurrence of convective core overshoot and rotation in intermediate-mass stars, and, by analysing a rich data sample of double-lined eclipsing binaries \citep[][]{2016A&A...592A..15C, 2017ApJ...849...18C, 2018ApJ...859..100C, 2019ApJ...876..134C}, they retrieved a calibrated core-overshooting efficiency parameter of $\lambda_\mathrm{ov}=0.4$. 
In the framework of the ballistic overshooting approach of \citet{bressan81}, this efficiency parameter corresponds to an averaged overshooting distance of $ d_{\rm ov}\sim 0.2~H_\mathrm{P}$, where $H_\mathrm{P}$ is the local pressure scale height. This value is expected to be suitable for all stars that already developed a fully convective core. In this regard, we divided the range of mass into three subintervals with different values of $\lambda_\mathrm{ov}$: Stars with $M_\mathrm{i}< M_\mathrm{O1}$ have radiative cores in the MS, and no convective overshoot is accordingly considered. Stars with $M_\mathrm{i}\geq M_\mathrm{O2}$ instead have a fully convective core, and the maximum overshooting efficiency ($\lambda_\mathrm{ov}=0.4$) is therefore used, and stars in the interval $M_\mathrm{O1}\leq M_\mathrm{i}\leq M_\mathrm{O2}$ have a developing convective core, for which we therefore apply an overshooting efficiency that linearly increases with the initial mass from $0$ to $0.4$. The adopted values of $M_\mathrm{O1}$ and $M_\mathrm{O2}$ at each metallicity are summarised in Table~\ref{tran_masses}. The same strategy was adopted for the convective envelope overshooting. The minimum value $\Lambda_\mathrm{e}=0.5$ was adopted at $M_\mathrm{i}< M_\mathrm{O1}$ and the maximum $\Lambda_\mathrm{e}=0.7$ was adopted at $M_\mathrm{i}\geq M_\mathrm{O2}$ \citep[see][]{1991A&A...244...95A,2018MNRAS.476..496F}. 

In \texttt{PARSEC v2.0}, the transport of angular momentum and chemical species is treated with a diffusive approach, as detailed in \paperone. Moreover, the shellular rotation scheme is adopted, with a constant angular velocity ($\Omega=$ constant) along the surface of each isobar \citep[see][]{2009pfer.book.....M}.

Rotation is parametrised by its rotation rate, $\omega=\Omega/\Omega_\mathrm{c}$, where $\Omega$ is the angular velocity, and $\Omega_\mathrm{c}=(2/3)^{3/2}\left(GM/R_\mathrm{pol}^3\right)^{1/2}$ is its critical value. In this expression, $G$ is the gravitational constant, and $M, R_\mathrm{pol}$ are the stellar mass and polar radius. In \texttt{PARSEC v2.0}, rotation sets in in a few models before the zero-age MS (ZAMS), assuming a solid-body profile with $\omega = \omega_\mathrm{i}$. 
From the ZAMS, the angular velocity evolves as the star evolves, following the angular momentum transport, the momentum loss due to stellar winds (or mechanical removal in case of critical rotation), and ensuring the conservation of angular momentum at each time-step. 

Moreover, we adopted the same method to impose a maximum initial rotation rate depending on the initial mass, following the observational results that low- and very low-mass stars rotate modestly or not at all \citep[see][]{2014ApJS..211...24M}. Therefore, models with \Mi $\leq$ $M_{\mathrm{O1}}$ were only computed with $\omega_\mathrm{i}=0.00$.
For models with a higher initial mass (\Mi $\geq$  $M_{\mathrm{O2}}$), we computed tracks with all values of the rotational rate. 
For masses with $M_{\mathrm{O1}}$ $\leq$ \Mi $<$ $M_{\mathrm{O2}}$, we only considered some values of the initial rotation rate up to a maximum value, which we computed as
\begin{equation}\label{omemax}
\omega_{\mathrm{i,max}}(M_\mathrm{i})= 0.99\left(\frac{M_\mathrm{i}-M_{\mathrm{O1}}}{M_{\mathrm{O2}}-M_{\mathrm{O1}}}\right).
\end{equation}

\begin{table}[!tbp]
\small
\caption{Initial masses at the transition between distinct overshooting prescriptions as a function of initial metallicity.}
\label{tran_masses}
\centering
\begin{tabular}{l | c c c c c c c }
\hline\hline
$Z$ &  $0.0001$ & $0.0002$ & $0.0005$ & $0.001$ & $0.002$ & $0.02$ & $0.03$ \\
\hline
$M_\mathrm{O1}$ & $0.96$ & $0.96$ & $0.98$ & $0.98$ & $1.02$ & $1.18$ & $1.16$ \\ 
$M_\mathrm{O2}$ & $1.26$ & $1.26$ & $1.28$ & $1.28$ & $1.32$ & $1.48$ & $1.46$ \\ 
\hline
\end{tabular}
\end{table}

\begin{table}[!tbp]
\small
\caption{ Initial mass value at the transition from low- to intermediate-mass stars as a function of the initial metallicity and rotation rate.}
\label{HeF_mass}
\centering
\begin{tabular}{l | c c c c c c c }
\hline\hline
$Z$ &  $0.0001$ & $0.0002$ & $0.0005$ & $0.001$ & $0.002$ & $0.02$ & $0.03$ \\
\hline
$\omega_\mathrm{i}$ & \multicolumn{6}{c}{$M_\mathrm{HeF}/\Msun$} \\
\hline
0.00 & $1.95$ & $1.95$ & $1.90$ & $1.85$ & $1.85$ & $2.05$ & $2.05$\\
0.30 & $2.15$ & $1.95$ & $1.90$ & $1.85$ & $1.85$ & $2.05$ & $2.05$\\
0.60 & $2.15$ & $1.95$ & $1.90$ & $1.90$ & $1.85$ & $2.05$ & $2.05$\\
0.80 & $1.95$ & $1.90$ & $1.90$ & $1.85$ & $1.85$ & $2.05$ & $2.05$\\
0.90 & $1.95$ & $1.90$ & $1.85$ & $1.80$ & $1.85$ & $2.05$ & $2.05$\\
0.95 & $1.90$ & $1.85$ & $1.80$ & $1.80$ & $1.80$ & $2.05$ & $2.05$\\
0.99 & $1.90$ & $1.85$ & $1.80$ & $1.75$ & $1.80$ & $2.05$ & $2.05$\\
\hline
\end{tabular}
\end{table}

\begin{figure*}
    \includegraphics[width=\textwidth]{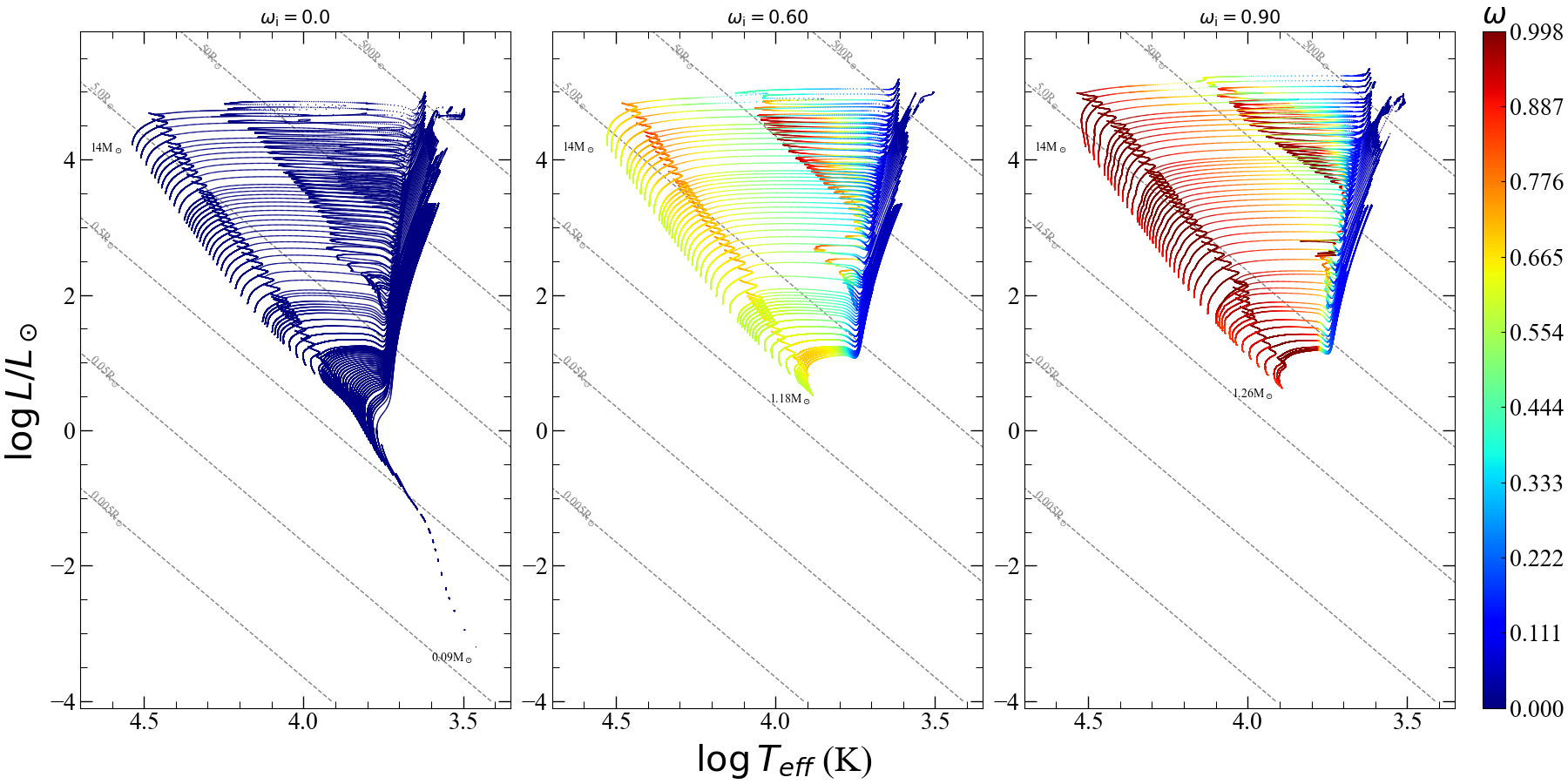}
    \caption{Hertzsprung–Russell diagram of stellar mass models in three sets of initial rotation rates: $\omega_\mathrm{i}=0.00,0.60$, and $0.90$ for a given metallicity $Z=0.001$, $Y=0.250$. The colour bar indicates the evolution of the rotation rate ($\omega$) of each single model. For the sake of clarity, the PMS evolution is cut off in this plot.}
    \label{HRD_Z0.001}
\end{figure*}

Mass loss becomes critically important in rotating stars because it is enhanced by rotational velocity \citep[][]{1986ApJ...311..701F,1993ApJ...409..429B}. In \texttt{PARSEC v2.0} models, mass loss is applied throughout the stellar evolution. The rotating mass-loss rate is enhanced by a factor from its non-rotating counterpart,
\begin{align}
    \dot{M}(\omega) = \dot{M}(\omega =0)\left(1-\frac{v}{v_{\mathrm{crit}}}\right)^{-0.43},
\end{align}
where $v$ is the surface tangential velocity, and $v_\mathrm{crit}$ is its critical value $v_\mathrm{crit}=(2GM/3R_\mathrm{pol})^{1/2}$. $\dot{M}(\omega =0)$ is the mass-loss rate of non-rotating models. Several mass-loss prescriptions were adopted in different mass ranges for $\dot{M}(\omega =0)$. In particular, for low-mass tracks, the empirical relation $\dot{M}\propto \eta M^{-1}L^{1.5}T_\mathrm{eff}^{-2}$ from \citet{1975MSRSL...8..369R,1977A&A....61..217R} was adopted, with the efficiency coefficient $\eta=0.2$ \citep{2012MNRAS.419.2077M}. For intermediate and massive tracks, the mass-loss rates from \citet{1988A&AS...72..259D} and \citet{2001A&A...369..574V} were adopted, respectively. They were both corrected for a factor that depended on the surface metallicity, that is, $\dot{M} \propto (Z/Z_\odot)^{0.85}$ \citep[see also][]{costa25}.

As in \paperone, we adopted the solar-scaled compositions of \citet{2011SoPh..268..255C}, except for the primordial elements. In short, the initial mass fraction of any metal $i$, $X_i$, scales linearly with $Z$, that is, $X_i = Z\frac{X_{i,\odot}}{Z_\odot}$, with $Z_\odot=0.01524$. 
We derived the initial He content from the enrichment law $Y=Y_p+\frac{\Delta Y}{\Delta Z}Z$, where $Y_p=0.2485$ is the primordial He content \citep{2011ApJS..192...18K}, and the helium-to-metal enrichment ratio $\Delta Y/\Delta Z=1.78$ was based on the solar calibration from \citet{2012MNRAS.427..127B}. 
On the other hand, for the three primordial elements,  $^3$He was taken as a fraction of $9\times 10^{-5}$ of the He content $Y$. For deuterium and $^7$Li, we distinguished between two cases: For $Z<0.001$, we adopt the Big Bang nucleosynthesis abundances (namely, $X_\mathrm{D}\approx 4.23\times 10^{-5}$ and $X_\mathrm{^7Li}\approx 2.72\times 10^{-9}$; see \citet{2012ApJ...744..158C}), whereas for $Z>Z_\odot$, we used the meteoritic values (namely, $X_\mathrm{D}\approx 2.12\times 10^{-5}$ and $X_\mathrm{^7Li}\approx 1.09\times 10^{-8}$). In the intermediate regime, the mass fractions were computed by a linear interpolation between these values. 
Grids of low-metallicity tracks computed with $\alpha$-enhanced mixtures \citep[][]{2018MNRAS.476..496F} are reserved for the coming releases.

We used an updated nuclear reaction network for a total of 72 different reactions that included the p-p chains,
the CNO tri-cycle, the Ne-Na and Mg-Al chains, the $^{12}$C, $^{16}$O
and $^{20}$Ne burning reactions, and the $\alpha$-capture reactions
up to $^{56}$Ni \citep[see][for details]{2012MNRAS.427..127B, 2018MNRAS.476..496F, 2021MNRAS.501.4514C}.  
The transport of energy by convection was described by the mixing length theory of \citet{1958ZA.....46..245B}, with the solar-calibrated mixing-length parameter $\alpha_\mathrm{MLT}=1.74$ from \citet{2012MNRAS.427..127B}. 

Finally, in agreement with the \texttt{PARSEC v2.0} tracks already released in \paperone, we computed seven initial rotation rates ( $\omega_\mathrm{i}$ = 0.0, 0.30, 0.60, 0.80, 0.90, 0.95, and 0.99). 
More detailed information is provided in the following section.

\section{Stellar tracks}
\label{sec:tracks}
All the tracks started at the pre-MS (PMS) phase and ended at a stage that depends on the initial mass: At an age that exceeds the Hubble time for very low masses, at the initial stages of the thermally pulsing asymptotic giant branch (TP-AGB) for low- and intermediate-mass stars, or at carbon exhaustion for more massive stars. 
We denote $M_{\rm HeF}$ as the value of initial mass of the most massive star undergoing the He-flash at the tip of the red giant branch (RGB) phase. This initial mass value also distinguishes between low- and intermediate-mass stars, and its dependence on the initial metallicity and rotational rate is summarised in Table ~\ref{HeF_mass}.

It should be noted that the $\omegai=0$ models with $M_\mathrm{i}\leq M_\mathrm{HeF}$ are interrupted at the beginning of the He flash, and they separately resumed at the zero-age horizontal branch with the same He core mass and surface chemical compositions as the last RGB model. The same models were then used for all $\omegai>0$ sequences of the same initial metallicity. 
This procedure is a first approximation that is needed to build the complete isochrones. 
On the other hand, this procedure ignores the changes in the surface chemical composition that are caused by rotation in the previous MS, at least for the core He-burning stars in the limited mass interval between $M_\mathrm{O1}$ and $M_\mathrm{HeF}$. The impact of rotation at this evolutionary stage will be addressed in coming work.

In the following, we discuss the impact of rotation on the photometric properties of stars (Sect.~\ref{sect_impact_rotation}) and on their surface chemical abundances (Sect.~\ref{surf_abund_sect}). Then, we compare our tracks with models from other available databases (Sect.~\ref{comp_other}).

\subsection{Impact of rotation on the stellar evolution and structure}\label{sect_impact_rotation}

Figure~\ref{HRD_Z0.001} shows the Hertzsprung–Russell diagram (HRD) of several mass models with three initial rotation rates. 
In rotating models, the local effective gravity is varied along the co-latitude angle because of the centrifugal force caused by rotation, and thus, the local effective temperature varies according to the von Zeipel theorem \citep[][]{1924MNRAS..84..684V}. 
Therefore, the value of effective temperature $T_\mathrm{eff}$ displayed in Fig.~\ref{HRD_Z0.001} is obtained as an average over the rotating isobaric surface. 
In the non-rotating set (left panel), we present all masses from $0.09-14~\Msun$. The tracks with an initial mass lower than $0.7~\Msun$ were adopted from \citet{2014MNRAS.444.2525C}. The intermediate rotational rate with $\omega_\mathrm{i}=0.60$ is shown in the central panel, 
where the lowest mass in this set is $1.18~\Msun$ as a result of the condition of Eq.~\ref{omemax}. Similarly, the right panel shows the set of fast-rotating models with $\omega_\mathrm{i}=0.90$, where the lowest mass is $1.26~\Msun$. 
The lowest masses in each set of rotation rates and metallicities are listed in Table~\ref{minimum_mass_rots}. 

\begin{table}[!tbp]
\caption{Lowest masses in each set of rotation rates and metallicities (including the sets from \paperone). 
}
\label{minimum_mass_rots}
\centering
\begin{tabular}{l | c c c c c c }
\hline\hline
$\omegai$ & $0.30$ & $0.60$ & $0.80$ & $0.90$ & $0.95$ & $0.99$ \\
\hline
$Z$ & \multicolumn{6}{c}{$M_\mathrm{i}/\Msun$} \\
\hline
$0.0001$ & $1.06$ & $1.16$ & $1.22$ & $1.24$ & $1.26$ & $1.26$ \\
$0.0002$ & $1.06$ & $1.16$ & $1.22$ & $1.24$ & $1.26$ & $1.26$ \\
$0.0005$ & $1.08$ & $1.18$ & $1.24$ & $1.26$ & $1.28$ & $1.28$ \\
$0.001$ & $1.08$ & $1.18$ & $1.24$ & $1.26$ & $1.28$ & $1.28$ \\
$0.002$ & $1.12$ & $1.22$ & $1.28$ & $1.30$ & $1.32$ & $1.32$ \\
$0.004$ & $1.16$ & $1.26$ & $1.32$ & $1.34$ & $1.36$ & $1.36$ \\
$0.006$ & $1.19$ & $1.29$ & $1.35$ & $1.37$ & $1.39$ & $1.39$ \\
$0.008$ & $1.24$ & $1.34$ & $1.40$ & $1.42$ & $1.44$ & $1.44$ \\
$0.01$ & $1.24$ & $1.34$ & $1.40$ & $1.42$ & $1.44$ & $1.44$ \\
$0.014$ & $1.26$ & $1.36$ & $1.42$ & $1.44$ & $1.46$ & $1.46$ \\
$0.017$ & $1.28$ & $1.38$ & $1.44$ & $1.46$ & $1.48$ & $1.48$ \\
$0.02$ & $1.28$ & $1.38$ & $1.44$ & $1.46$ & $1.48$ & $1.48$ \\
$0.03$ & $1.26$ & $1.36$ & $1.42$ & $1.44$ & $1.46$ & $1.46$ \\
\hline
\end{tabular}
\end{table}

As the stars evolve along the MS, the surface angular velocity is increased because of angular momentum transport, and so is the rotation rate. It reaches the highest value around the turn-off region. 
After the MS, the core contracts and the envelope significantly expands, which significantly decreases the surface angular velocity due to the angular momentum conservation. 
As low-mass stars evolve further along the RGB and increase their radius, the surface velocity drops rapidly while the angular velocity of the inner core increases \citep[see][]{2014ApJ...788...93C,2022A&A...665A.126N}. 

\begin{figure}
    \centering
    \includegraphics[width=\linewidth]{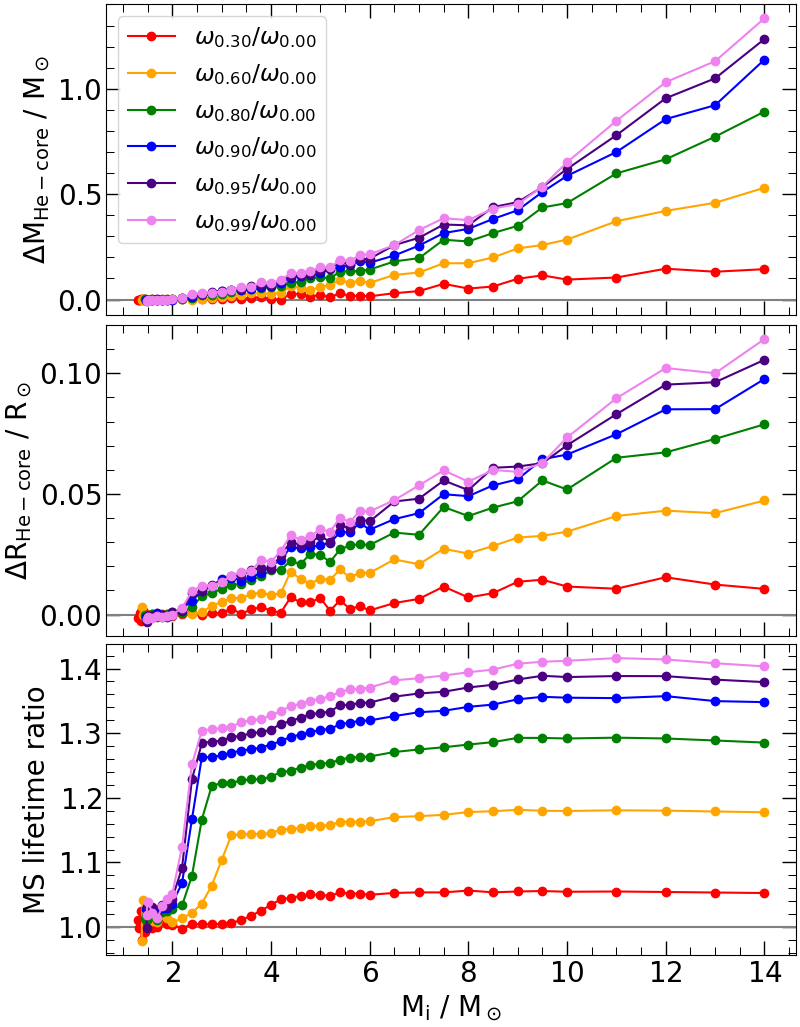}
    \caption[]{Difference between rotating models and their non-rotating counterparts in He-core mass (top panel), He-core radius (middle panel) at the TAMS ($X_\mathrm{c}\sim 10^{-7}$), and the MS lifetime ratios (bottom panel). Models from low- to intermediate-mass are shown here with an initial metallicity $Z=0.02$.
    }
    \label{He_core}
\end{figure}

During the MS, the intermediate-mass rotating stars become more luminous than their non-rotating counterparts because the core mass increases significantly through rotational mixing. It also increases the core radius and MS lifetime. For example, in the case of $M_\mathrm{i}=6 \Msun$, the track with $\omegai=0.99$ has a MS lifetime that is longer by $1.37$ times than for its non-rotating counterpart. It simultaneously increases its core mass and radius by approximately $0.22 \Msun$ and $0.043 \Rsun$. 
This increase is illustrated in Fig.~\ref{He_core} for models with a varying initial mass and \omegai. 
This has consequences for their post-MS phases as well, although their surface velocity evolves similarly to that of low-mass stars. 
In the He-burning phase (or blue loop), the contraction of their envelope later increases $\omega$. 

\begin{figure*}
    \includegraphics[width=\textwidth]{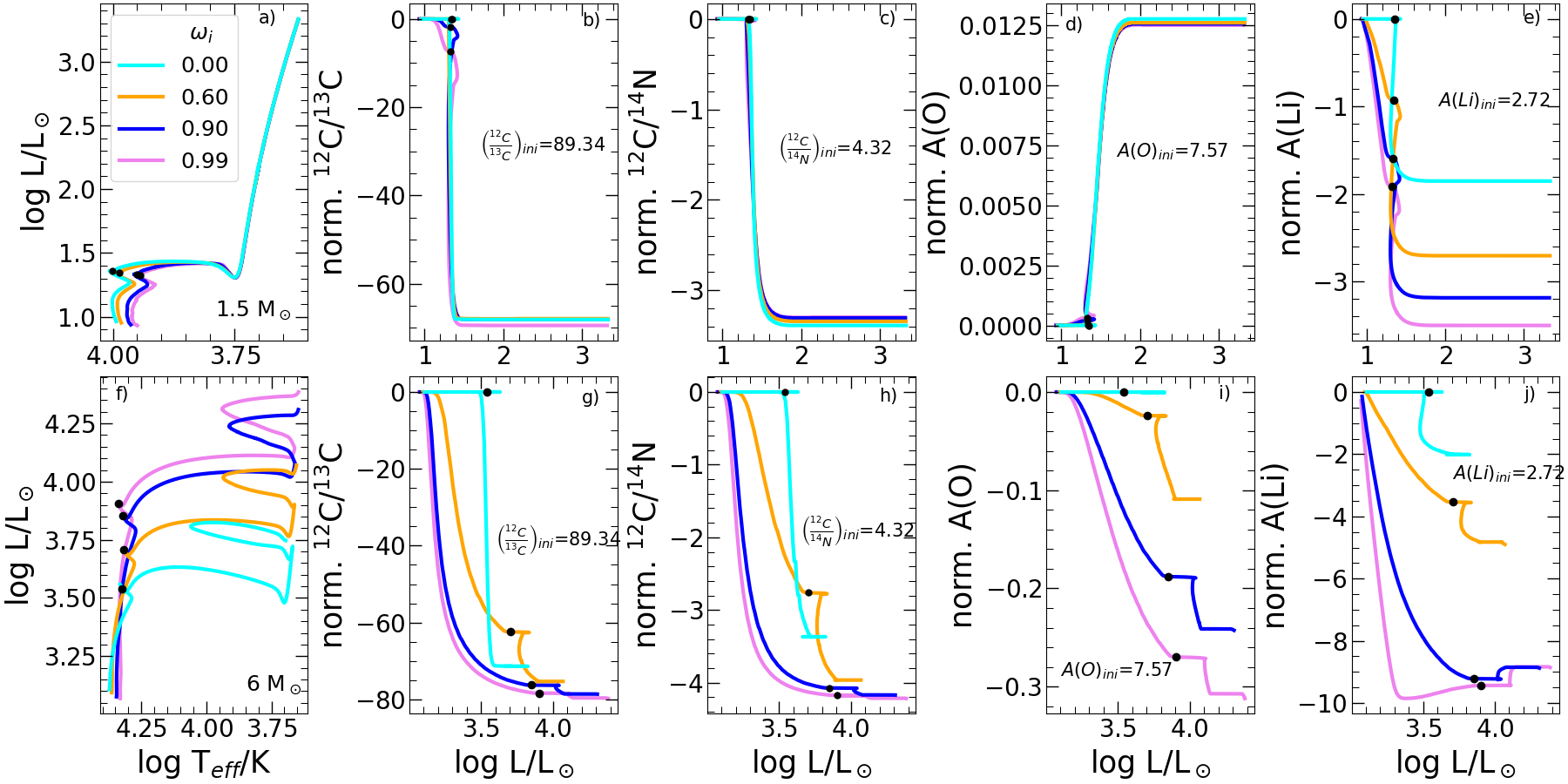}
    \caption{Impact of rotation on the HRD and several surface abundances normalised to their initial values. Models with a mass of $1.5~\Msun$ are shown in the top row (panels a-e), and models with a mass of $6~\Msun$ are shown in the second row (panels f-j). Different initial rotation rates are shown by different colours. The first column shows the HRD of the selected models. The second and third columns show the C-isotope ratio and the C/N ratio, respectively, with respect to the initial values. That is, we plot the quantity $(X/Y)-(X/Y)_\mathrm{ini}$, where $X/Y= n_X/n_Y$ is number density ratio of species $X$ and $Y$, and the $(X/Y)_\mathrm{ini}$ value is specified in the plot. The abundances of $^{16}$O and $^7$Li are shown in the fourth and fifth columns, respectively. In the latter cases, we plot the quantity $A(X)-A(X)_\mathrm{ini}$, where $A(X)=\log(n_X/n_\mathrm{H})+12$ is the abundance of element $X$, and $A(X)_\mathrm{ini}$ is its initial value, as specified in each panel. $n_\mathrm{H}$ is the hydrogen density number. The models are shown from the ZAMS to the RGB tip for the $1.5~\Msun$ model, and at the end of the He-burning phase for the $6~\Msun$. The black dot marks the TAMS.}
    \label{isotopes_variation}
\end{figure*}

\subsection{Impact of rotation on surface abundances}\label{surf_abund_sect}

In the surface chemical abundances of stars, stellar internal mixing processes can be proved, especially today, when large spectroscopic surveys provide accurate abundances of several elements, for example, the GALactic Archaeology with HERMES \citep[GALAH,][]{2019A&A...624A..19B}, Apache Point Observatory Galactic Evolution Experiment \citep[APOGEE,][]{2022ApJS..259...35A}, and Gaia-ESO \citep{2022A&A...666A.120G} for hundreds of thousands of stars. Our \texttt{PARSEC v2.0} evolutionary tracks provide predictions for the surface abundances of several elements (from $^1$H to $^{60}$Zn) along the evolution. Five of these elements are retained in the isochrone tables\footnote{We recall that surface abundances of other elements and isotopes (e.g., $^{18}$O, $^{13}$C, $^{7}$Li, etc.) can be added to the isochrones as well, upon reasonable request to the main authors.}, namely $^1$H, $^4$He, $^{12}$C, $^{14}$N, and $^{16}$O. 

Panels b to j of Fig.~\ref{isotopes_variation} show the evolution of several surface abundances and their ratios with respect to the initially assumed values for a few selected elements and stars with and without rotation of initial masses of $1.5~\Msun$ and $6~\Msun$. 
In particular, Fig.~\ref{isotopes_variation} (b-g, c-h) shows the variation in the C-isotopes and the CN ratios. In standard non-rotating models (cyan lines), the ratios remain the same as their initial values during the MS evolution because there is no active mixing process. 
The first change in the surface abundances occurs at the beginning of the red giant branch when the downward extension of the convective envelope causes the first dredge-up (1DU). 
As a consequence, the nuclear burned products $\mathrm{^{13}C}$ and $\mathrm{^{14}N}$ are brought up to the surface, while $\mathrm{^{12}C}$ is depleted \citep[see also][]{1967ARA&A...5..571I,2017hsn..book..461K}.

When rotation is considered, rotational mixing is established. In the case of intermediate-mass stars where the CNO cycle is the main burning channel (e.g. for the $6~\Msun$ star in Fig.~\ref{isotopes_variation}), a significant decrease in the $^{12}$C$/^{13}$C and $^{12}$C$/^{14}$N ratios from their initial values can be seen during the MS phase. The later evolution shows further depletion mainly due to the 1DU. In general, the decrease is stronger for a higher initial rotational rate. 
The situation is different in the case of low-mass stars, where the pp-chains are the main burning channels during the MS, and rotational mixing becomes insignificant: Overall, rotation has little impact on the variation in these ratios. This also holds true for the oxygen abundance (Fig.~\ref{isotopes_variation}d). 
Instead, rotational mixing in intermediate-mass stars depletes surface oxygen far below the values predicted by the non-rotating 1DU models (Fig.~\ref{isotopes_variation}i).
More extreme rotation rates cause even stronger depletions. This might provide a good explanation for the chemically anomalous stars observed in many open clusters \citep[e.g.][]{2018PASJ...70...91T,2021A&A...656A.155L,2022AJ....164..255C}.

On the other hand, Fig.~\ref{isotopes_variation}e and j show the variation in surface lithium. A significant amount of lithium can be transferred from the envelope to the inner layers through rotational mixing, where it is eventually destroyed at temperatures of $\sim 2.6\times 10^6$~K. 
The efficiency of Li depletion increases with the initial rotation rate, mass, and eventually, the metallicity, as we show below. The need for rotation to explain the observed abundances of lithium in supergiant stars has been suggested variously before \citep[e.g.][]{2012MNRAS.427...11L, 2021A&A...651A..84M, 2022ApJ...931...61F}. 
The discrepancy between the primordial value \citep[A(Li)=2.69;][]{2014JCAP...10..050C} and the Spite plateau \citep[A(Li)=2.3;][]{1982A&A...115..357S} measured from dwarf-MS low-mass metal-poor stars, which is commonly known as the cosmological lithium problem, requires additional mixing mechanisms in the early evolution up to the terminal-age mains sequence (TAMS) to explain it \citep[see][for references]{2005ApJ...619..538R,2015MNRAS.452.3256F,2025A&A...696A.136N}. We clarify that in the current release, \texttt{PARSEC v2.0} uses the standard schemes for additional mixing, that is, classical atomic diffusion \citep[][]{2012MNRAS.427..127B} and a fixed value for the efficiency of envelope overshooting. 
It is therefore reasonable to expect that additional physics needs to be implemented in our models in order to accurately reproduce the Li abundance that is observed in low-mass metal-poor stars.

\begin{figure}
    \centering
    \includegraphics[width=\linewidth]{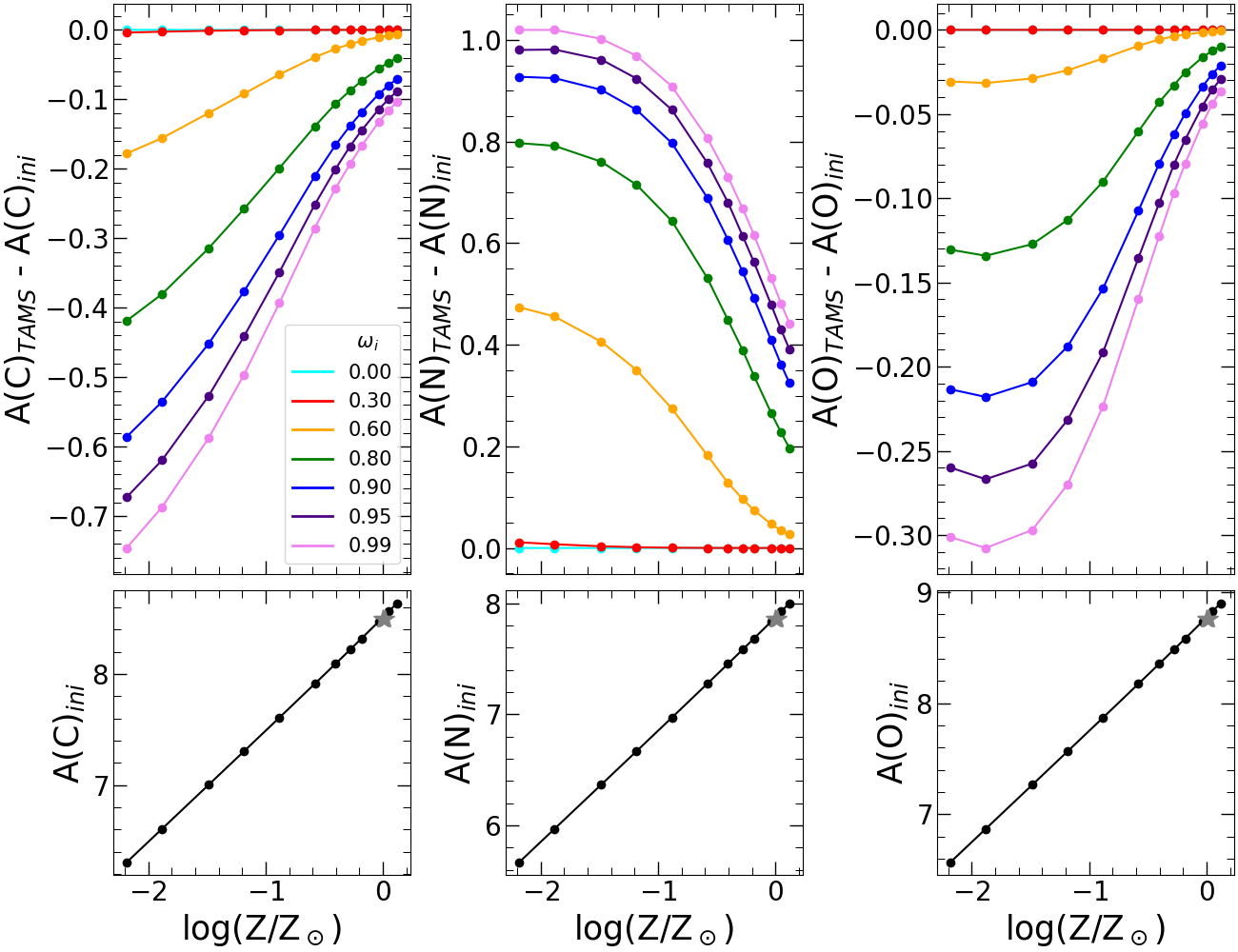}
    \caption{Upper panel: Difference of the surface CNO abundances between the TAMS and initial value for $6~\Msun$ stellar tracks with different metallicities and various initial rotation rates. Bottom panel: Initial values of the CNO abundances. The grey star indicates the initial abundances at solar metallicity.}
    \label{DX_CNO}
\end{figure}

In Fig.~\ref{DX_CNO} we show the impact of rotation on the CNO abundances of a $6~\Msun$ track for different metallicities as a representative of intermediate-mass stars. The bottom panels show the initial abundance of CNO isotopes of our tracks with $13$ different metallicities. 
The upper panels show the difference in the surface CNO abundances between the TAMS and the initial values. 
The non-rotating models (cyan) show no changes at all metallicities. On the other hand, in the presence of rotational mixing, C and O are depleted and the N abundance is enhanced. 
More interestingly, Fig.~\ref{DX_CNO} implies that the effect of rotational mixing on the CNO abundances also depends on metallicity, in addition to stellar mass and rotational rate: Its efficiency increases significantly towards lower metallicities because the convective core is more extended. 

\begin{figure}
    \centering
    \includegraphics[width=\linewidth]{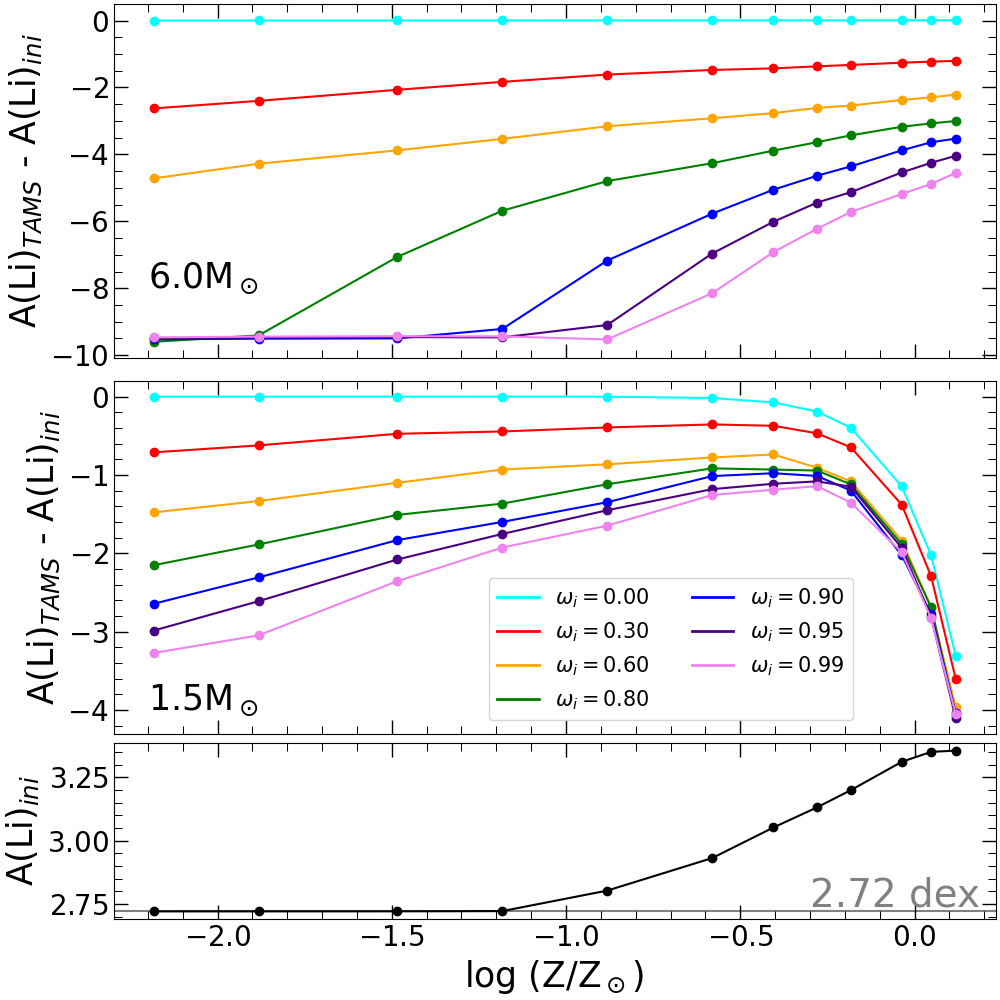}
    \caption{Same as Fig.~\ref{DX_CNO}, but for lithium. The upper panel shows the abundance difference for $6~\Msun$ stars, the middle panel shows it for $1.5~\Msun$ stars, and the bottom panel is the initial $^7$Li-abundance at a given metallicity.}
    \label{DX_Li7}
\end{figure}

Similar results are found for the case of lithium, which is shown in the upper panel of Fig.~\ref{DX_Li7}, for the $6~\Msun$ model. The depletion of lithium caused by rotation ultimately depends on the rotation rate, mass, and metallicity. 
The middle panel of Fig.~\ref{DX_Li7} also shows the depletion of lithium at $\log (Z/Z_\odot)\geq -0.6$ for the non-rotating $1.5~\Msun$ stars. 
This can be understood by the fact that the convenctive cores of low-mass stars with a higher metallicity are more extended. The temperature at the bottom of the envelopes is already sufficient to destroy lithium during the early evolution. This process is clearly aided by convective envelope overshoot in the models. A dedicated investigation of the lithium evolution in the PMS of low-mass stars will be pursued in a forthcoming work.

\subsection{Comparison to \texttt{GENEC} and \texttt{MIST}}\label{comp_other}
\begin{figure}
    \centering
    \includegraphics[width=\linewidth]{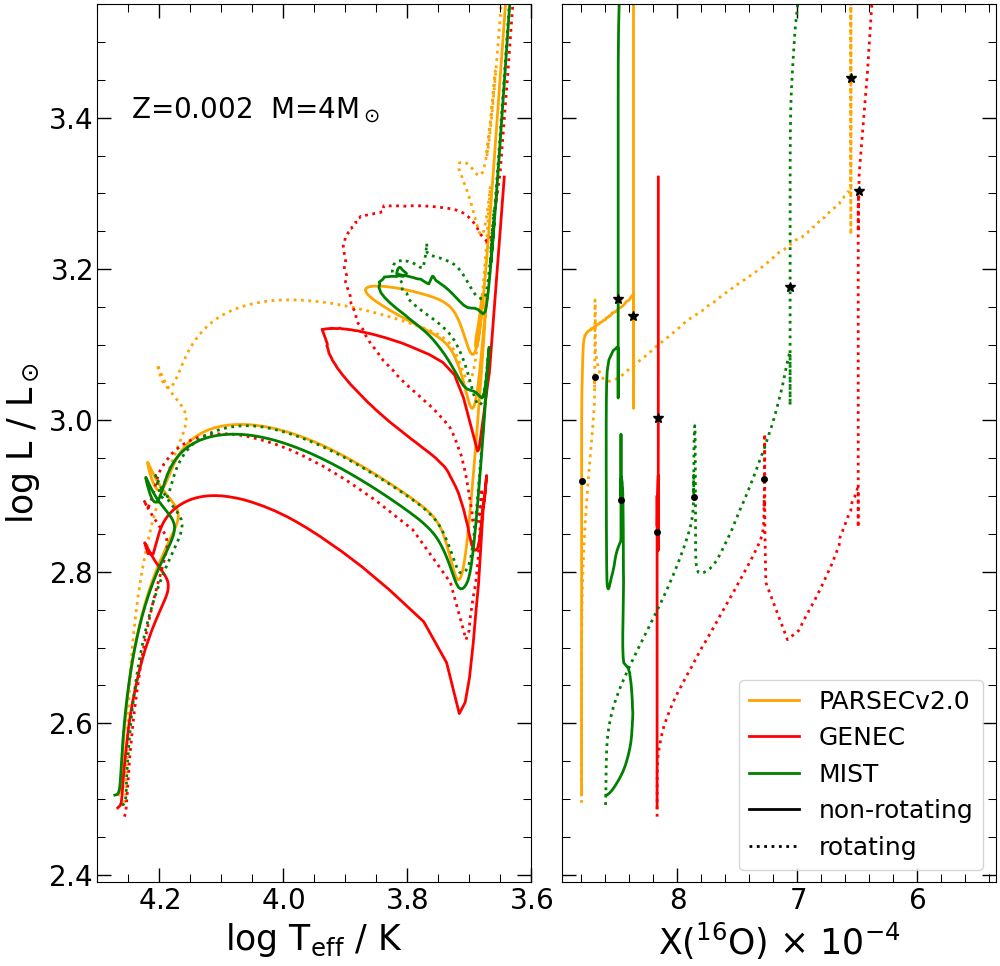}
    \caption{Comparison between the stellar tracks in terms of HRD evolution (left panel) and the surface oxygen mass fraction evolution (right panel). 
    All tracks have an initial metallicity $Z=0.002$ and initial mass $M_\mathrm{i}=4~\Msun$. Non-rotating tracks are shown as solid lines, and rotating tracks are shown as dotted lines. 
    The colour-code indicates the tracks obtained from different databases. Rotating \texttt{PARSEC v2.0} and \texttt{GENEC} tracks are with $\omegai=0.59$, and \texttt{MIST} is with $\nu/\nu_\mathrm{crit}=0.4$. 
    All stellar tracks begin from the ZAMS. The black dots mark the TAMS, and black stars mark the end of the He-burning phase.}
    \label{O_GENEC_comp}
\end{figure}

The differences between \texttt{PARSEC v2.0} and its previous version (PARSEC v1.2S) were already extensively discussed in \paperone. In this subsection, we compare our stellar tracks with the \texttt{GENEC} tracks \citep{2013A&A...558A.103G} and from the \texttt{MIST} database \citep[][]{2016ApJ...823..102C}. 
\texttt{GENEC} computed tracks with the initial rotation rate defined by the tangential velocities, $v/v_\mathrm{crit}=0.40$, which corresponds to $\Omega/\Omega_\mathrm{crit}=0.59$ in terms of angular velocities. 
Therefore, we computed a few additional tracks ($Z=0.002$, $M=4~\Msun$) with this initial value of $\omega_\mathrm{i}=0.59$ for the sake of comparison. 
Fig.~\ref{O_GENEC_comp} indicates that \texttt{PARSEC v2.0} tracks are systematically brighter, have less extended blue loops, and have a milder depletion of the oxygen abundance than \texttt{GENEC} tracks.

We note that the input physics in the \texttt{PARSEC v2.0}, \texttt{GENEC}, and \texttt{MIST} tracks differ widely. Table~\ref{blue_loop_compare} presents some of the differences regarding the mixing length parameter, the $\mathrm{^{12}C(\alpha,\gamma)^{16}O}$, the $\mathrm{^{14}N(p,\gamma)^{15}O}$ nuclear reaction rates, and the envelope-overshooting efficiency. 
In particular, \texttt{GENEC} adopts an overshoot distance of $0.1 H_P$ from the Schwarzschild border of the core, while \texttt{PARSEC v2.0} applies a value of $0.2 H_P$ (or, equivalently, a $\lambda_\mathrm{ov}=0.4$ across the border; see \citealt{bressan81}). This difference immediately leads to a higher He-core mass in \texttt{PARSEC v2.0}, and thus, to a more luminous track after the MS. 
This also affects the MS lifetime. For example, the non-rotating $4~\Msun$ model shows a difference of $14$ Myr between the two codes. Namely, \texttt{PARSEC v2.0} predicts $138$ Myr, and \texttt{GENEC} predicts $124$ Myr. We also verified the non-rotating $4~\Msun$, $Z=0.002$ computed by \texttt{STAREVOL} code \citep[][]{2012A&A...543A.108L}, with which the MS lifetime of this track is similar to the prediction of \texttt{GENEC}, that is, $126$ Myr. 
Additionally, all the above ingredients directly affect the extent and luminosity of the blue-loop feature of intermediate-mass stars, as shown by \citet{2014MNRAS.445.4287T}.

We also show in Fig.~\ref{O_GENEC_comp} the interpolated tracks, taken from \texttt{MIST} database, with an initial metallicity $Z=0.002$ and mass $M_\mathrm{i}=4~\Msun$. The HRD evolution of the non-rotating \texttt{PARSEC v2.0} track is similar to that of \texttt{MIST}. This is mainly because of the equivalent core-overshooting efficiency parameter adopted in \texttt{PARSEC v2.0} and \texttt{MIST} tracks. 
Despite the different overshooting implementation (ballistic step for \texttt{PARSEC v2.0} and exponential decay for \texttt{MIST}), they are about equivalent to a step-overshooting distance of $0.2~H_P$ \citep[see][]{2010ApJ...718.1378M, 2016ApJ...823..102C}. 
They use different criteria, Schwarzschild versus Ledoux, to define the convective regions, which leads to a slight difference in the advanced phases. The \texttt{MIST} track predicts a similar MS lifetime. In particular, the non-rotating $4~\Msun$ \texttt{MIST} track gives $137$ Myr. 

\begin{table}[!tbp]
\caption{Input parameters adopted in three stellar evolutionary codes.
}
\label{blue_loop_compare}
\centering
\begin{tabular}{l c c c c }
\hline\hline
 & \texttt{GENEC}  & \texttt{PARSEC v2.0} & \texttt{MIST} \\
 \hline
$\alpha_\mathrm{MLT}$ &  $1.6$ & $1.74$ & $1.82$ \\

$\mathrm{^{12}C(\alpha,\gamma)^{16}O}$ &  KU02 & CHW0 & BU96 \\

$\mathrm{^{14}N(p,\gamma)^{15}O}$ & MU08 & IM05 & IM05 \\

$\Lambda_\mathrm{e}$ (in $H_P$) & no & $0.7$ (step) & $0.0174$ (exp.) \\

$d_\mathrm{ov}$ (in $H_P$) & $0.1$ (step) & $0.2$ (step) & $0.016$ (exp.)\\
solar-mixtures & AS05 & CA11 & AS09 \\
initial $\nu$ (km/s) & $230$ & $226$ & $250$ \\
\hline
\end{tabular}
\tablefoot{KU02: \citet{2002ApJ...567..643K}; 
CHW0: \citet{Cyburt2012};
MU08: \citet{PhysRevC.78.015804}; IM05: \citet{2005EPJA...25..455I}; BU96: \citet{1997ApJ...479L.153B}; AS05: \citet{2005ASPC..336...25A}; CA11: \citet{2011SoPh..268..255C}; AS09: \citet{2009ARA&A..47..481A}.
}
\end{table}

For the comparison of rotating tracks, we selected models with similar initial tangential velocity at the ZAMS. 
Despite the initial similar configuration, the evolution proceeds differently, not only due to the different non-rotating physical parameters, but also to the different implementation and treatment of angular momentum transport. 
In particular, \texttt{GENEC} models treat meridional circulation as an advection process and the shear instability as a diffusion process. 
In \texttt{PARSEC v2.0}, the diffusion scheme is instead used to treat both rotational instabilities. 
\texttt{MIST} model treats rotation still differently by taking five different rotationally induced instabilities into account \citep[][]{2013ApJS..208....4P, 2016ApJ...823..102C}. 
These differences directly lead to variations in the distribution of chemical elements along the evolution between different tracks. 
Regardless of the differences in the initial mass fractions of the adopted chemical mixtures (see Table \ref{blue_loop_compare}), the impact of rotational mixing can be seen by the amount of surface oxygen that is depleted throughout the evolution. 
For example, in the case shown here, the TAMS of the \texttt{PARSEC v2.0} track decreases modestly $\Delta X_O\approx 1\times 10^{-5}$ from its initial value. On the other hand, \texttt{GENEC} tracks deplete it more strongly, $\Delta X_O\approx 9\times 10^{-5}$, which can only be due to the different efficiencies of rotational mixing. 
Similar to \texttt{GENEC}, the rotating \texttt{MIST} track with $\nu/\nu_\mathrm{crit}=0.4$ shows $\Delta X_O\approx 7\times 10^{-5}$. 

After the TAMS, when the stars begin 1DU, together with the effect of rotational mixing, the rotating tracks show a significant depletion of surface oxygen. This is seen in all \texttt{PARSEC v2.0}, \texttt{GENEC}, and \texttt{MIST} rotating tracks. The same trends are seen for the depletion of $^{12}$C and the enhancement of $^{14}$N.

\section{Isochrones}
\label{sec:isochrones}

In \paperone, isochrones were derived for the same seven values of $\omega_\mathrm{i}$ at which evolutionary tracks were computed. This involved interpolating in the plane of initial mass and age defined by every grid of evolutionary tracks of a given $\omega_\mathrm{i}$. Interpolation between tracks of different metallicities was also allowed. 

Since then, we improved the interpolation algorithms of the TRIdimensional modeL of thE GALaxy code \citep[\texttt{TRILEGAL},][]{2005A&A...436..895G,2017ApJ...835...77M} so as to also allow for the interpolation between grids with different $\omega_\mathrm{i}$. The interpolation method is essentially the same as was described by \citet{bertelli08} to produce isochrones at different values of metallicity and helium content. This means that isochrones can now be produced for any $\omega_\mathrm{i}$ value in the interval from 0.00 to 0.99. This new interpolation scheme is now available at the CMD v3.8 web interface\footnote{\url{http://stev.oapd.inaf.it/cgi-bin/cmd_3.8}}.

\begin{figure}
    \centering
    \includegraphics[width=\linewidth]{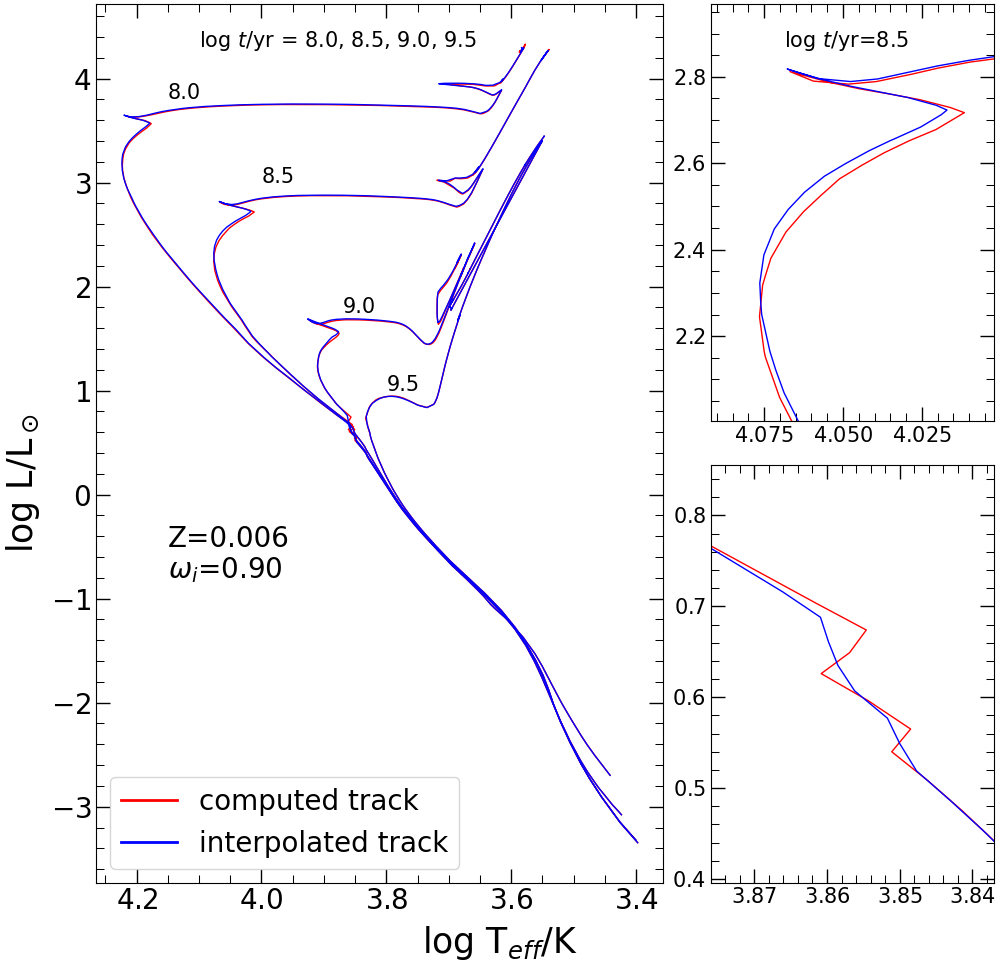}
    \caption{Comparison of theoretical isochrones produced by the computed stellar tracks (red) and the interpolation algorithm (blue). The isochrones are for an initial metallicity of $Z=0.006$, $\omegai=0.90$, and the ages are indicated in the plot. The right panels are zoomed in to the turn-off region (upper panel) and the lower-MS part (bottom panel) of the isochrones of $\log t/\mathrm{yr}=8.5$.}
    \label{interp_check}
\end{figure}

To test this new interpolation algorithm, we produced a number of isochrones of a given $\omegai=0.90$ and $Z=0.006$ that were computed 1) with the actual set of computed tracks, and 2) by the interpolation between two neighbouring values of $\omegai=0.80$ and $0.95$. Fig.~\ref{interp_check} shows the produced isochrones by the two methods for ages $\log t/\mathrm{yr}=8.0-9.5$. The isochrones produced by the interpolation algorithm are very similar to those produced directly from the $\omegai=0.90$ tracks. Small differences can be found just in the turn-off region, with the level of $\Delta T_\mathrm{eff}=0.006$ dex (upper right panel), and at the transition mass at which rotation is assumed to start being efficient (bottom right panel). It is expected that the typical errors caused by the interpolation in $\omegai$ are smaller than these indicative examples.

With the new tracks, we computed the corresponding isochrones, from which we also derived the absolute magnitudes by applying the bolometric corrections (BCs) by \citet{2019MNRAS.488..696G}. 
These BCs were suitably computed for rotating stars as a function of the stellar inclination $i$. They were made available for the passbands of many different photometric systems \citep[see][]{2019A&A...632A.105C}. 

We note that the inclination-dependent BCs described by  \citet{2019MNRAS.488..696G} cannot be computed for the coolest stars because current libraries of model atmospheres are limited. This is less of a problem in our case because $\omega$ of initially fast-rotating stars is reduced to values of $\sim\!0.1$ when the stars reach a mean effective temperature cooler than $\sim\!5000$ K. These stars deviate very little from spherical symmetry, and they can be described by a single $\Teff$ value. Therefore, for $\Teff < 5130$ K, we adopted the BCs derived from non-rotating models (i.e. derived from plane-parallel model atmospheres, as in \citealt{girardi02}). In the interval from 5370 to 5130 K, we smoothly interpolated between the BCs derived from rotating stars and those derived from non-rotating stars.

Finally, a brief explanation of the quantities presented in the isochrone tables can be found in a dedicated document\footnote{\url{https://stev.oapd.inaf.it/cmd_3.8/help.html}}. For the quantities presented in the track tables, we refer to anther dedicated document\footnote{\url{https://stev.oapd.inaf.it/PARSEC/Database/PARSECv2.0_VMS/readme.html}}.

\section{Comparison with observations}
\label{sec:summary}

In order to verify the quality of our models, we compared them with observations of the young cluster NGC~6067. The age of this cluster was estimated between 50 and 150 Myr \citep[][]{1962MNRAS.124..445T,1993MNRAS.260..915S,2013Ap&SS.347...61M}, and the metallicity varies from solar to supersolar value. For instance, \citet{1995AJ....110.2813P} derived a mean metallicity of $\mathrm{[Fe/H]=-0.01\pm 0.07}$ dex, while \citet{2017MNRAS.469.1330A} suggested a mean value of $0.19\pm 0.05$ dex. The metallicity of the renowned Cepheid V340~Nor in this cluster is estimated to be a near-solar value, for which different works agree well (e.g. \citet{2017MNRAS.469.1330A} claimed $\mathrm{[Fe/H]=0.09\pm 0.11}$ dex, and \citet{2014A&A...566A..37G} claimed $\mathrm{[Fe/H]=0.07 \pm 0.1}$ dex). The reddening and distance modulus of NGC~6067 are also well constrained in the literature. Recent studies reported a reddening $E(B-V)\approx 0.34 \pm 0.04$~mag \citep[e.g.][]{2007ApJ...671.1640A,2020MNRAS.496.4701J,2022MNRAS.509.1664J}. For the true distance modulus, \citet{2020A&A...640A...1C} reported a value of ${(m-M)_0}=11.37\pm 0.2$~mag, while \citet{2021MNRAS.503.1864P} claimed a value of $11.74\pm 0.26$~mag, and \citet{2022MNRAS.509.1664J} provided an intermediate value of $11.62\pm 0.15$~mag. Within the uncertainties, they tend to agree with each other.

\begin{figure}
    \centering
    \includegraphics[width=\linewidth]{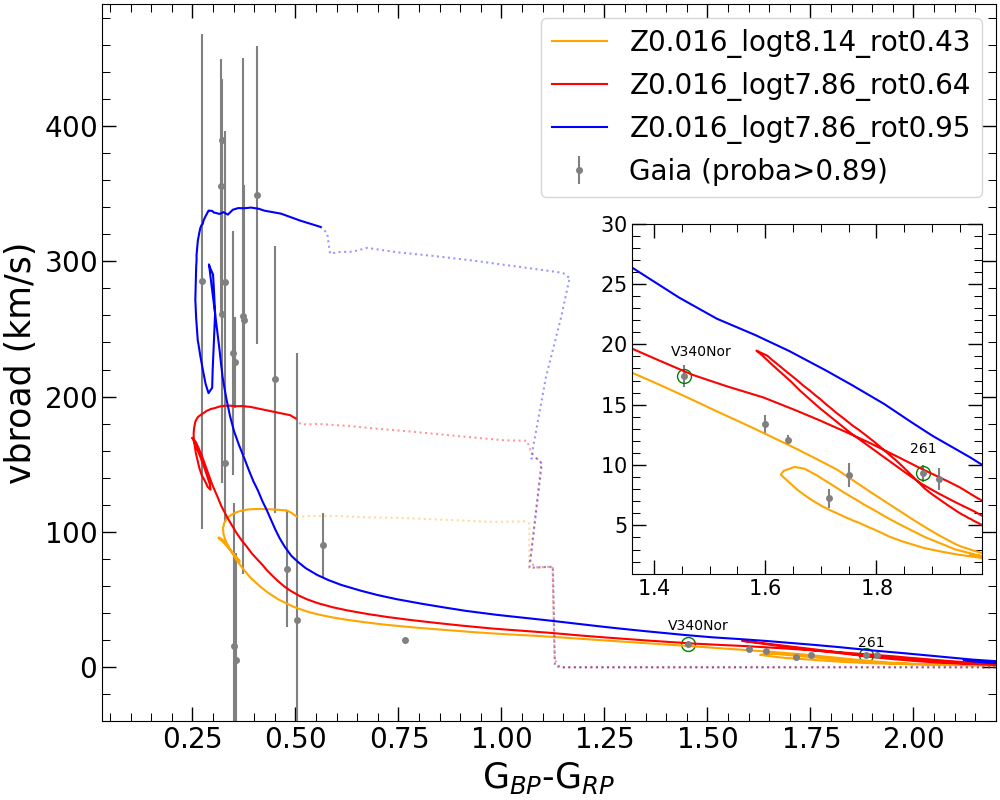}
    \caption{Broadening velocity of 25 star members of NGC 6067. The surface tangential velocity from the isochrone models is shown with an inclination angle $i=90^o$ ($i=0^o$ stars have a similar range in colour and negligible \texttt{vbroad}). The solid line represents the upper MS evolution and after ($G_\mathrm{mag}<14$ mag). The dotted lines represent the lower MS part of the isochrones ($G_\mathrm{mag}>14$ mag). The inset panel zooms into the cool-stars region.}
    \label{vbroad_ngc6067}
\end{figure}

\begin{figure}
    \centering
    \includegraphics[width=\linewidth]{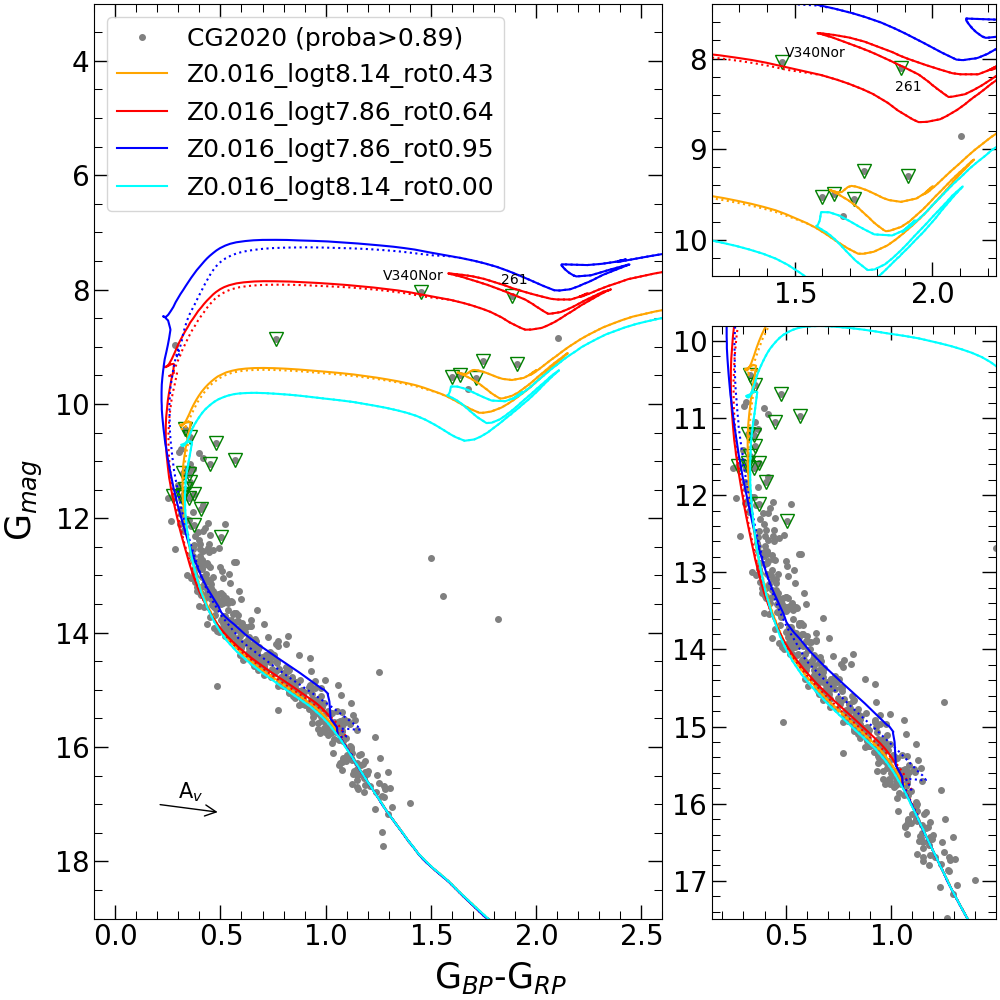}
    \caption{CMD of NGC 6067. The data were adopted from the Gaia DR2 catalogue \citep[][]{2020A&A...633A..99C}, and stars with a membership probability $\geq 0.89$ are shown. Stars with \texttt{vbroad} are shown as green triangles. The isochrones with the adopted metallicity ($Z=0.016$), ages ($\log t/\mathrm{yr}=7.86$ and $8.14$), and initial rotational rates ($\omega_\mathrm{i}=0.43$, $0.64$ and $0.95$) are plotted as solid ($i=0^\circ$) and dotted lines ($i=90^\circ$).}
    \label{cmd_ngc6067_gaia}
\end{figure}

Moreover, the recent Gaia DR3 data contain the broadening velocity (\texttt{vbroad}) derived with the Radial Velocity Spectrometer (RVS) for more than three million bright ($G<12$ mag) targets \citep[][]{2023A&A...674A...1G}. This broadening velocity includes the contributions of the projected rotational velocity and the macroturbulent velocity, as well as other effects. For rotating stars, \texttt{vbroad} was shown to be an excellent proxy for the projected rotational velocity, $v\sin i$ \citep[see also][]{2024MNRAS.532.1547C}. 

We compared the observed \texttt{vbroad} with the tangential velocity predicted in our models \citep[see also][]{2024MNRAS.532.1547C}. Before we start, we note that the tangential velocity should be compared with the upper values of \texttt{vbroad} because the observed velocity is reduced by the (unknown) projection factor $\sin i$. 

Figure~\ref{vbroad_ngc6067} shows the variation in \texttt{vbroad} versus the Gaia colour of stars in the cluster NGC~6067. The data were retrieved from the Gaia DR3 archive\footnote{\url{https://gea.esac.esa.int/archive/}} \citep[][]{2023A&A...674A...8F}. 
In total, we found that only $53$ stars from the NGC~6067 sample have available \texttt{vbroad} values because the catalogue is currently limited to $G_\mathrm{RVS}<12$ mag. By selecting stars with a high membership probability of only ($\texttt{proba} \geq 0.89$), we reduced the sample to $25$ stars. 
Among these, stars at the MS turn-off show a wide range of \texttt{vbroad} values spanning from a few dozen km/s up to the extremely fast at $\sim 400$ km/s. On the other hand, cool stars on the red side have significantly lower \texttt{vbroad} values (lower than $\sim20$ km/s), as expected from the significant expansion of their envelopes, which rapidly decreases the surface rotational velocity. 
We clarify that the Gaia $vbroad<10$ km/s is unreliable because the measurements were limited \citep[see][]{2023A&A...674A...8F}. Nonetheless, the cepheid V340~Nor, which is very well studied and was used as a benchmark for many calibrations, is presented in the sample with $vbroad \sim 17.4\pm 9.3$ km/s. In order to determine the location of V340~Nor in the plot, we cross-matched its (RA, Dec) from the sample of \citet{2017MNRAS.469.1330A} with the Gaia catalogue.  

In Fig.~\ref{vbroad_ngc6067} we plot the tangential velocities predicted from our models are plotted as solid lines. Our extremely fast rotating model with $\omega_\mathrm{i}\approx 0.95$ reproduces the extremely high \texttt{vbroad} in the turn-off region very well. Stars with a milder \texttt{vbroad} at the turn-off are either stars with similar \omegai\ observed at small inclination, or stars with $\omegai\simeq0.64$ observed at high inclination. The same $\omegai\simeq0.64$ values suffice to explain the \texttt{vbroad} of the cool stars, including 340\_Nor, although models with higher \omegai\ and a low inclination cannot be excluded. On the other hand, a significant group of four cool stars could be explained with a lower rotation rate, namely $\omegai=0.43$, implying that they might be observed at high inclination.

The CMD fit in the Gaia passbands is shown in Fig.~\ref{cmd_ngc6067_gaia} with the adopted data from \citet{2020A&A...633A..99C}, where only
members with the probability $\geq 0.89$ are selected, together with the $25$ stars whose published \texttt{vbroad} are shown as green triangles. 
The results of \citet{2017MNRAS.469.1330A} indicated a spread in metallicity within the observed stars in their sample. In particular, 4 of the $13$ cool stars have $\mathrm{[Fe/H]}\leq 0.1$ dex, including the cepheid V340~Nor. The other stars have a mean value of about $\mathrm{[Fe/H]}\approx 0.2$ dex. By using the Padova isochrones \citep[][]{2000A&AS..141..371G} where rotation was not yet implemented, \citet{2017MNRAS.469.1330A} found no conclusive evidence of multiple populations in this cluster. 
In this work, we revisited the CMD of NGC 6067 with our rotating models and explored the multiple populations in this cluster.

For this purpose, an intermediate true distance modulus of $(m-M)_0=11.50$ mag was adopted from the literature \citep[see][]{2020A&A...640A...1C,2021MNRAS.503.1864P,2022MNRAS.509.1664J}. The extinction $A_V=1.10$ mag was selected so as to fit the position of MS stars ($G_\mathrm{mag}>10$). Although the metallicity spread found by \citet{2017MNRAS.469.1330A} is still debated, we chose a metallicity slightly above the solar value for our isochrones ($Z=0.016$) and investigated whether a higher metallicity is needed. 
Our isochrone with an age of $\log (t/\mathrm{yr})=8.14$ ($\sim 138$ Myr), and $\omega_\mathrm{i}=0.43$ fits the cool stars best (with $G_\mathrm{mag}>9$), which are assumed to be in the core He-burning phase, and with the MS part of the cluster. For the most luminous stars (Cepheid V340~Nor and the K2 type 261 star), our isochrone with the same metallicity but with a younger age, $\log t=7.86$ ($\sim 72$ Myr) and higher $\omega_\mathrm{i}=0.64$ fits the data well. As mentioned above (Fig.~\ref{vbroad_ngc6067}), the choice of these rotating models explains the observed \texttt{vbroad} of these cool stars well. Models with extremely high rotation or non-rotating models, on the other hand, fail to reproduce the CMD of this cluster. In particular, the $\omega_\mathrm{i}=0.95$ model with $\log (t/\mathrm{yr})=7.86$ is too bright to reproduce the location of the luminous stars. Similarly, with the same choice of parameters, the non-rotating model fails to reproduce the He-burning stars.

\begin{figure}
    \centering
    \includegraphics[width=\linewidth]{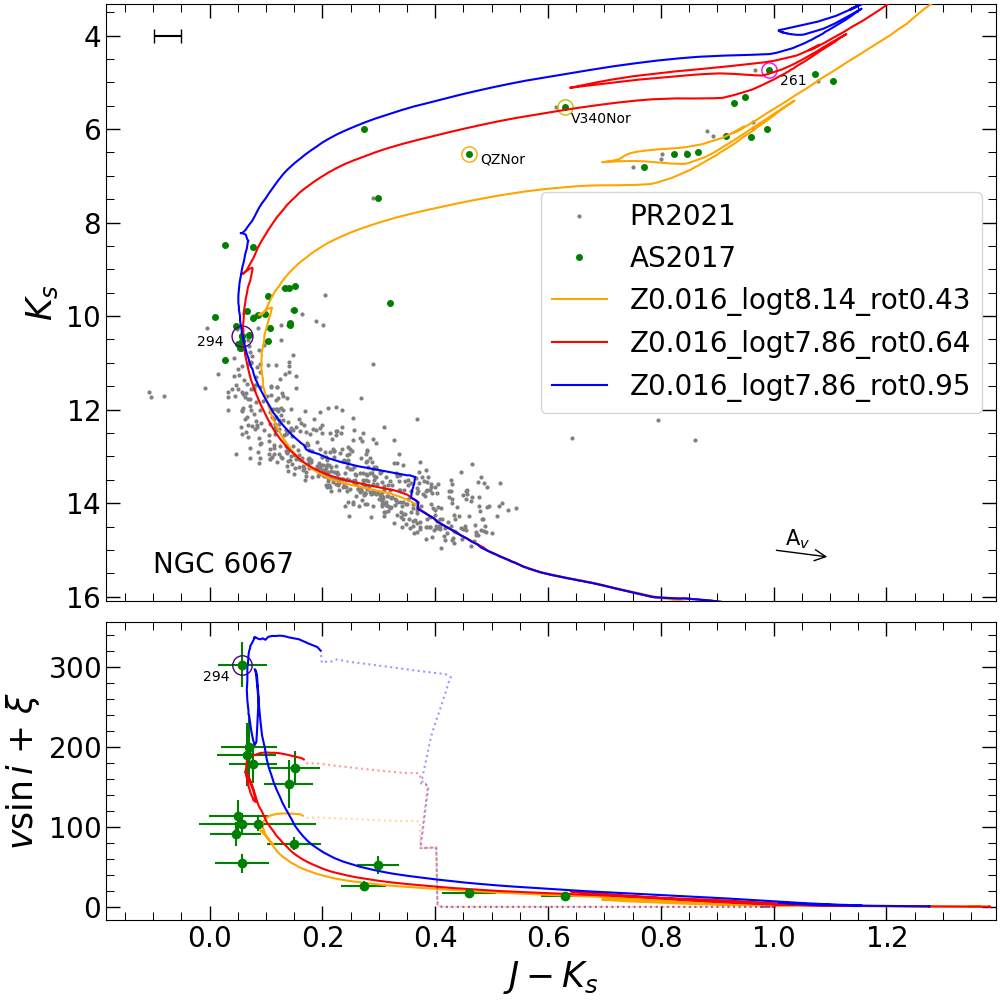}
    \caption{Top panel: CMD of NGC 6067 in Vista photometry. The green dots are data adopted from \citet{2017MNRAS.469.1330A}, and the grey dots are data adopted from \citet{2021MNRAS.503.1864P}. The superimposed isochrones are models with the same parameters as in Figs.~\ref{vbroad_ngc6067} and \ref{cmd_ngc6067_gaia}. Bottom panel: Rotational velocity of stars from the \citet{2017MNRAS.469.1330A} catalogue, superimposed with the predictions from isochrone models (solid lines show the evolution with $K_\mathrm{S}<13$ mag, and dotted lines show $K_\mathrm{S}>13$ mag). Only stars with precise $v\sin i$ and $\xi$ are presented in this plot.}
    \label{ngc6067_CMD_vista}
\end{figure}

As a result, the Gaia CMD fits show that no higher metallicity is required for NGC 6067. To further confirm the fits above, we performed the CMD fit based on the VISTA photometry system, as shown in Fig.~\ref{ngc6067_CMD_vista}. The upper panel shows the observed CMD for a total of $42$ stars from \citet{2017MNRAS.469.1330A} (green dots), together with the sample of \citet{2021MNRAS.503.1864P} in the background (grey dots). The best-fit models from Fig.~\ref{cmd_ngc6067_gaia} were adopted in this plot. The isochrones in the VISTA CMD agree well with our findings with the Gaia passbands.

The location of the cepheid QZ~Nor in Fig.~\ref{ngc6067_CMD_vista} is not reproduced very well by the isochrones. Nonetheless, it is unclear whether QZ~Nor is a member of NGC~6067 \citep[see][]{2021MNRAS.503.1864P}, and we thus chose not to rely on QZ~Nor for our discussion of the model fits above.

The bottom panel of Fig.~\ref{ngc6067_CMD_vista} shows the sum of the projected rotational velocity ($v\sin i$) and the macroturbulent velocity ($\xi$) of the stars in the sample of \citet{2017MNRAS.469.1330A}. We only selected non-binary stars with precise measured values in both velocities for the plot. The sum velocity of cepheid V340~Nor is $13\pm 5.6$ km/s. Our rotating models fit its velocity very well. Most stars in the MS turn-off are reproduced by our best-fit models, with a velocity below $200$ km/s. 
The B-type star with an emission line (Be-star) 294 is an exception. It has $v\sin i=279.6\pm 9.1$ km/s and $\xi=23.7 \pm 19.4$ km/s, which are well reproduced by our extremely fast-rotating models ($\omega_\mathrm{i}=0.95$) in general. Figs.~\ref{vbroad_ngc6067} and \ref{cmd_ngc6067_gaia} show, however, that this model fails to reproduce the observed CMD of the cluster in the Gaia and VISTA filters.

Similarly, \citet{2012A&A...538A.110M} also found that Be-stars are extremely fast-rotating stars. They are surrounded by a gaseous circumstellar environment that might be the origin of their fast-rotating behaviour \citep[see also][]{2006A&A...459..137R}. Moreover, \citet{2013A&A...553A..25G} suggested a mass and/or $T_\mathrm{eff}$ dependence on the rotational rate in order to explain the Be star phenomenon. 
A detailed model for the Be star phenomenon is beyond the scope of this paper, however.

\begin{figure}
    \centering
    \includegraphics[width=\linewidth]{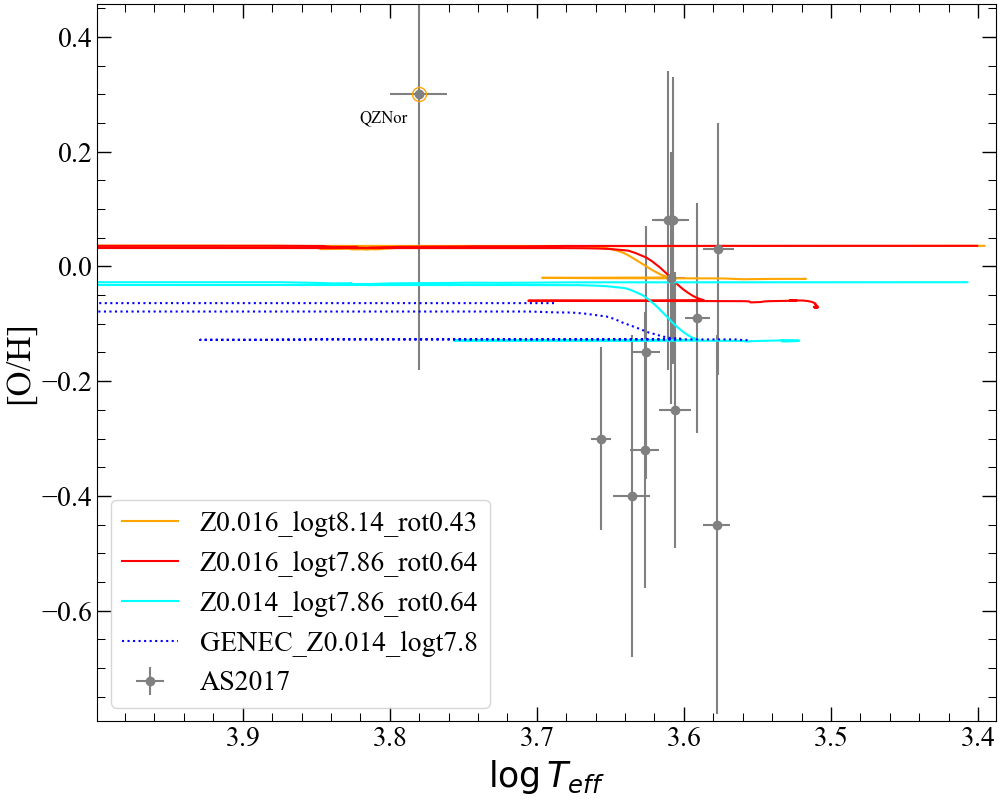}
    \caption{Oxygen-hydrogen ratio of cool stars by \citet{2017MNRAS.469.1330A}. The prediction from best-fit models is shown as solid lines. The prediction from the \texttt{GENEC} model ($Z=0.014$, $\log t=7.8$, $\omega_\mathrm{i}=0.59$) is plotted for comparison.}
    \label{ngc6067_[OH]_logTeff}
\end{figure}

Furthermore, \citet{2017MNRAS.469.1330A} reported the surface abundances of many cool stars in their sample. We paid particular attention to the oxygen abundance of these stars because our isochrone tables include the variation of three isotopes, $^{12}$C, $^{14}$N, and $^{16}$O. Fig.~\ref{ngc6067_[OH]_logTeff} showed their observed [O/H] values, together with the prediction from our best-fit models. First of all, the measured [O/H] are clearly spread: Six stars have nearly solar values, but five stars are strongly depleted in oxygen. The isochrones with different ages indicate a very modest change in [O/H], while rotation enhances the depletion of the O abundance, as expected. Our best-fit $\omega_\mathrm{i}=0.64$ model is not enough to explain the four stars that are significantly depleted, however, even with their $1\sigma$ uncertainty. The model with a lower metallicity might be able to explain the strongly [O/H] depleted stars. The rotating \texttt{GENEC} model \citep[][]{2012A&A...537A.146E} with $Z=0.014$ provides a similar prediction. The prediction from both models suggests that a lower metallicity is required in order to reproduce the significant depletion [O/H]. On the other hand, the result from the CMD fits above shows no clear evidence of a metallicity spread in this cluster. We also recall, however, that the data show large error bars. This makes it difficult to reach any firm conclusion on the properties of the cluster stars. 
A more thorough selection of their chemical abundances together with a more detailed analysis are required. We will address this in forthcoming works. 

The discussion above illustrates the general difficulties of simultaneously interpreting current data for open clusters on photometry, projected velocities, and abundance variations. It also show that large sets of isochrones which include the effect of rotation are clearly needed in the era of Gaia data and of large spectroscopic surveys. 

\section{Summary and conclusion}

We presented the new calculation of stellar evolutionary tracks and isochrones for seven metallicities to complement the first release of \texttt{PARSEC v2.0} in \paperone. In total, we covered the metallicity range from $Z=0.0001$ to $0.03$. In order to be homogeneously consistent with the first release, we adopted the same input physics as \paperone for our calculations. We covered the mass range from $\sim 0.7-14~\Msun$, with seven initial rotation rates spanning from $\omegai=0.00$ to $0.99$ that depended on the initial masses. The details of the input physics are described in Sect.~\ref{sec:inputphysics}, and the impact of rotation on the stellar structure and evolution is studied in Sect.~\ref{sec:tracks}.

In this second release, we include the surface abundances of several more isotopes (from $^1$H to $^{60}$Zn) in the stellar track table. We simultaneously present five of them ($^1$H, $^4$He, $^{12}$C, $^{14}$N, $^{16}$O) in the isochrone table, with the possibility of adding more isotopes upon request. We also upgrade the isochrone web-interface, in which the isochrones with an intermediate rotation rate between the computed values can now be archived by an interpolation scheme that is described in Sect.~\ref{sec:isochrones}.

We verified the quality of our new computed models by comparisons with the observed data of the open cluster NGC~6067 in terms of broadening velocity, photometric colour-magnitude diagram, and oxygen abundances. The results indicate that rotating models are crucial for explaining the extremely fast-rotating stars in the turn-off region of the cluster. On the other hand, our rotating models suggest that at least two populations are harboured in the cluster, with a dispersion in age and initial rotation rate. The populations have the same metallicity. In contrast, the observed oxygen abundance of cool-star members requires a lower metallicity. The large uncertainty from the observed [O/H] prevented us from finding conclusive evidence in this regard. 

Finally, to reiterate, all the stellar evolutionary tracks we discussed, including the previous database presented in \paperone, are available at a dedicated website\footnote{\url{http://stev.oapd.inaf.it/PARSEC/}}. The corresponding isochrones with the new interpolation in the initial rotational rate can be obtained at another dedicated website\footnote{\url{http://stev.oapd.inaf.it/cgi-bin/cmd}}.

\begin{acknowledgements}
We thank the referee (Devesh Nandal) for the thorough and constructive comments/suggestions. 
This project has received funding from the European Union’s Horizon 2020 research and innovation programme under grant agreement No 101008324 (ChETEC-INFRA). We also acknowledge the financial support from INAF Theory Grant 2022. We acknowledge the Italian Ministerial grant PRIN2022, ``Radiative opacities for astrophysical applications'', no. 2022NEXMP8. GC acknowledges partial financial support from European Union—Next Generation EU, Mission 4, Component 2, CUP: C93C24004920006, project ‘FIRES'. The contribution by MT and GP is funded by the European Union – NextGenerationEU and by the University of Padua under the 2023 STARS Grants@Unipd programme (``CONVERGENCE: CONstraining the Variability of Evolved Red Giants for ENhancing the Comprehension of Exoplanets''). 
CTN acknowledges the support by INAF Mini grant 2024, ``GALoMS – Galactic Archaeology for Low Mass Stars''. 
AJK acknowledges support by the Swedish National Space Agency (SNSA). 
GE acknowledges the contribution of the Next Generation EU funds within the National Recovery and Resilience Plan (PNRR), Mission 4 - Education and Research, Component 2 - From Research to Business (M4C2), Investment Line 3.1 - Strengthening and creation of Research Infrastructures, Project IR0000034 – ``STILES - Strengthening the Italian Leadership in ELT and SKA''. 
YC acknowledges the Natural Science Research Project of Anhui Educational Committee No. 2024AH050049, National Natural Science Foundation of China (NSFC) No. 12003001, the Anhui Project (Z010118169)
\end{acknowledgements}

\bibliographystyle{aa}
\bibliography{references} 

\makeatletter

\def\thebiblio#1{
 \list{}{\usecounter{dummy}
         \labelwidth\z@
         \leftmargin 1.5em
         \itemsep \z@
         \itemindent-\leftmargin}
 \reset@font\small
 \parindent\z@
 \parskip\z@ plus .1pt\relax
 \def\newblock{\hskip .11em plus .33em minus .07em}
 \sloppy\clubpenalty4000\widowpenalty4000
 \sfcode`\.=1000\relax
}
\let\endthebiblio=\endlist
\makeatother

\label{lastpage}

\end{document}